\documentclass[amsmath,amssymb,prc,final,showpacs,twocolumn]{revtex4-1}
 
\usepackage{bm,hyphenat,xspace}
\usepackage{graphicx,epsfig}

\newcommand {\mbf}[1]{{\mathbf{#1}}}

\newcommand{\fm}{\;\mathrm{fm}}

\newcommand{\Hh}{{}^3\mathrm{H}}

\begin{document}

\title {Three-neutron resonance study using transition operators}
  
\author{A.~Deltuva} 
\email{arnoldas.deltuva@tfai.vu.lt}
\affiliation
{Institute of Theoretical Physics and Astronomy, 
Vilnius University, Saul\.etekio al. 3, LT-10257 Vilnius, Lithuania
}

\received{December 27, 2017} 
\pacs{21.30.-x, 21.45.-v}

\begin{abstract}
\begin{description}
\item[Background]
Existing bound-state type calculations of three-neutron resonances
yield contradicting results.
\item[Purpose]
A direct study of the three-neutron continuum using rigorous 
scattering equations  with realistic potentials
and search for possible resonances is aimed.
\item[Methods]
Faddeev-type integral equations  for  three-neutron 
transition operators are solved 
 in the momentum-space partial-wave framework. The evolution 
of resonances is studied by enhancing the strength of the
two-neutron interaction in partial waves with nonzero orbital
momentum. 
\item[Results]
Calculated  three-neutron  transition operators exhibit resonant behavior
for sufficiently large enhancement factors; pole trajectories
in the complex-energy energy plane are extracted from
their energy dependence. However, the  resonant behavior completely 
disappears for the physical interaction strength. 
\item[Conclusions]
There are no physically observable three-neutron resonant states 
consistent with presently accepted interaction models.
\end{description}
\end{abstract}

 \maketitle

\section{Introduction \label{sec:intro}}

After a possible experimental observation of the four-neutron 
($4n$) resonance
\cite{PhysRevLett.116.052501}, 
a number of theoretical studies of multineutron systems emerged 
\cite{PhysRevC.93.044004,PhysRevLett.117.182502,PhysRevLett.118.232501,PhysRevLett.119.032501}. 
Their conclusions are, however, quite contradicting,
even for the simplest three-neutron ($3n$) system. While 
earlier studies \cite{PhysRevC.66.054001,lazauskas:3n}
based on the complex-scaled Faddeev equation
found no $3n$ resonances that could be physically observable,
a recent work \cite{PhysRevLett.118.232501} predicted a $3n$ resonance 
about 1 MeV above the threshold that should be
potentially measurable. However, the latter studies relied
on bound-state type calculations with extrapolation  to the continuum.
To shed more light on the possible existence and
observability of the $3n$ resonance,  a direct study of
the $3n$ continuum using rigorous scattering equations  
is the aim of the present work. 
The integral equation formulation of the scattering theory for transition
operators realized in the momentum-space partial-wave framework
will be used. An important advantage 
of the direct continuum approach is its ability to estimate not only the 
resonance position but also its effect on scattering amplitudes
that lead to observables in collision processes.

Section \ref{sec:eq} describes three-particle scattering equations
and some details of calculations whereas Sec. III reports
results for a number of interaction models.
The conclusions are presented in Sec. IV.

\section{Theory \label{sec:eq}}

Faddeev equations for three-particle transition operators
in the version proposed by Alt, Grassberger, and Sandhas (AGS)
\cite{alt:67a} 
have been extensively used for the description
of the nucleon-deuteron scattering
\cite{koike:87b,gloeckle:96a,deltuva:03a}. Using the
odd-man-out notation, the multichannel
transition operators $U_{\beta\alpha}$ 
 satisfy the integral equations
\begin{subequations} \label{eq:AGS}
\begin{equation} \label{eq:AGS1}
U_{\beta\alpha} =  \bar{\delta}_{\beta\alpha} G_0^{-1} + 
\sum_{\gamma} \bar{\delta}_{\beta\gamma} t_\gamma G_0 \, U_{\gamma\alpha}
\end{equation}
or, equivalently,
\begin{equation} \label{eq:AGS2}
U_{\beta\alpha} =  \bar{\delta}_{\beta\alpha} G_0^{-1} + 
\sum_{\gamma}  U_{\beta\gamma} G_0 \,  t_\gamma 
\bar{\delta}_{\gamma\alpha}.
\end{equation}
\end{subequations}
Here $\bar{\delta}_{\beta\alpha} = 1 - {\delta}_{\beta\alpha}$, 
$G_0 = (E+i0-H_0)^{-1}$ is the free resolvent at 
 the available  three-particle energy $E$ in the center-of-mass (c.m.) 
frame,  $H_0$ is the free Hamiltonian for the relative motion,
and 
\begin{equation} \label{eq:t}
t_\gamma = v_\gamma  + v_\gamma G_0 t_\gamma 
\end{equation}
is the two-particle transition for the pair $\gamma$ with
  $v_\gamma$ being the corresponding potential.
The sums over the spectator (pair) label $\gamma$ 
in Eqs.~(\ref{eq:AGS}) run from 1 to 3, thereby coupling 
only components corresponding to spectator + pair partitions.
In the $3n$ system there are no bound pairs, the only 
possible reaction is the
 elastic scattering of three free particles
($3\to3$ process) whose operator can be obtained from  $U_{\beta\alpha}$ 
with $\alpha,\beta = 1,2,3$ via the quadrature
\begin{equation} \label{eq:AGS00}
U_{00} =  \sum_{\alpha} t_\alpha  + 
\sum_{\beta \alpha} t_\beta \, G_0 \, U_{\beta\alpha} \, G_0 \, t_\alpha.
\end{equation}

For identical
particles the system (\ref{eq:AGS1}) reduces to a single equation for the
symmetrized transition operator
\begin{equation} \label{eq:AGSs}
U =  P G_0^{-1} + P t\, G_0 \, U
\end{equation}
with the two-particle transition operator $t$ for the
representative pair 1 
and  $P =  P_{12}\, P_{23} + P_{13}\, P_{23}$,
$P_{\beta\alpha}$ being
the permutation operator of particles $\alpha$ and $\beta$;
the basis states must be antisymmetric only under the exchange
of the neutrons within the pair. 
It is convenient to introduce an auxiliary Faddeev operator
$T = t G_0 U G_0 t$ obeying the integral equation
\begin{equation} \label{eq:AGSt}
T =  t G_0 P t + t G_0 P T
\end{equation}
since  it is more directly related to the $3\to3$ 
transition operator
\begin{equation} \label{eq:AGS00s}
U_{00} =  (1+P)t(1+P)  + (1+P)T(1+P).
\end{equation}
The first term describes the two-neutron ($nn$) 
scattering with the remaining one
being a spectator and therefore does not correspond to a genuine
three-particle process. Thus, for the investigation of the $3n$
 dynamics and possible resonances one should study the
behavior of  operators $U$ or $T$, not $U_{00}$.

The AGS equations (\ref{eq:AGSs}) and (\ref{eq:AGSt})
 are solved in the momentum-space.
After the partial-wave decomposition
 they become a system of integral equations 
with two continuous variables, the magnitudes of the 
 Jacobi momenta  for the pair $\mbf{p} = (\mbf{k}_2 - \mbf{k}_3)/2$
and for the spectator  $\mbf{q} = (2\mbf{k}_1 - \mbf{k}_2 - \mbf{k}_3)/3$
where $\mbf{k}_\alpha$ are the individual momenta.
 The associated orbital angular momenta $L$ and $l$
together with neutron spins $s_\alpha=\frac12$, through
intermediate angular momenta $s$, $j$, and $S_q$,  are coupled 
to the  total angular momentum  $J$ with the projection  $M$, 
resulting in the basis states $| pq \nu \rangle
= | pq  (l \{[L (s_2s_3)s]j \, s_1\}S_q ) \,{JM} \rangle$
with the total parity $\Pi = (-1)^{L+l}$
where $\nu$ abbreviates all discrete quantum numbers.
Due to the  antisymmetry condition only  even $(L + s)$ states are considered.
The results are well converged by including two-neutron states
with total angular momentum $j<3$, i.e.,
 ${}^1S_0$, ${}^3P_0$, ${}^3P_1$,  ${}^3P_2$,  ${}^3F_2$, and ${}^1D_2$ 
in the usual spectroscopic ${}^{2s+1}L_j$  notation. 

The numerical solution technique, including also the treatment
of kernel singularities, is taken over from
Ref.~\cite{deltuva:phd}. However, when studying the
resonant behavior of transition operators one has to avoid 
special kinematic situations where already  the on-shell driving term 
$\langle p'q'\nu'| t G_0 P t | pq \nu \rangle$ in Eq.~(\ref{eq:AGSt})
 becomes singular due to kinematic reasons and leads to
divergences in $\langle p'q'\nu'| T | pq \nu \rangle$
 for particular combinations of initial and final momenta,
i.e., for
${p'}^2 + 3{q'}^2/4 = {p}^2 + 3{q}^2/4 = mE$
and ${q'}^2+ {q}^2 \pm qq' = mE$,  with $m$ being the neutron mass.
In fact, such a situation corresponds to a free (on-shell) scattering
of two-neutrons followed by a free scattering of two-neutrons
within another pair, and therefore may be considered as a non-genuine 
three-particle reaction. In contrast, the $3n$ resonance, 
corresponding to the pole of
$T$ and $U$  in the complex-energy plane, manifests itself in
\emph{all} matrix elements of these transition operators,
also fully off-shell. In the vicinity of the pole $E_r -i \Gamma/2$ 
the transition operator in the corresponding $J^\Pi$ state
 can be expanded in series
\begin{equation} \label{eq:Upole}
 T_{J^\Pi} =  \sum_{n=-1}^\infty \tilde{T}^{(n)}_{J^\Pi} \, (E-E_r + i \Gamma/2 )^{n}
\end{equation}
and well approximated by few lowest terms while
higher-order terms yield negligible contribution.

Such a resonant behavior (or its absence) will be demonstrated
using three types of initial and final states differing in their
momentum distributions:

(1) q-state: on-shell state with $p=0$ and $q=q_m=\sqrt{4mE/3}$;
vanishing momentum $p$ implies ${}^1S_0$ state
for the pair while $l$ takes  one of $J \pm \frac12$ values
consistent with total parity;

(2) p-state: on-shell state with $p=p_m=\sqrt{mE}$ and $q=0$;
the second condition implies $l=0$ for the spectator;

(3) off-shell state: Gaussian momentum distribution
of 1 $\fm^{-1}$ width for the pair and  momentum
$q=\sqrt{4m(E+\epsilon_{\rm off})/3}$ for the spectator.

These state types in the following will be indicated  by 
superscripts  ``q'', ``p'', and ``off'',   respectively,
e.g., the  state with $p=0$, $q=q_m$, $L=s=j=0$
will be abbreviated by  ${}^1S_0^q$.

\section{Results \label{sec:res}}

A number of force models are used  for the present study
of the $3n$ system:

(1) A realistic high-precision charge-dependent Bonn
(CD Bonn) potential \cite{machleidt:01a} that 
was not applied to the $3n$ system so far.

(2) A realistic Reid93 potential \cite{stoks:94a}
already used in Ref.~\cite{lazauskas:3n} where
no physically observable $3n$ resonance was found.

(3) Chiral effective field theory ($\chi$EFT) potential at next-to-leading
 order (NLO) ~\cite{PhysRevLett.115.122301},
an improved version of the local NLO potential
used in Ref.~\cite{PhysRevLett.118.232501} that
predicts a $3n$ resonance about 1 MeV above threshold.
The central value for the regulator $R = 1.0$ fm is taken.
The three-nucleon force (3NF) appears only at higher order 
but its contribution in multineutron systems is
insignificant \cite{PhysRevLett.118.232501}.

(4) A realistic Argonne V18 potential \cite{wiringa:95a} 
whose low- and high-momentum components are partially
decoupled by the similarity 
renormalization group (SRG) transformation \cite{bogner:07b,bogner:07c}. 
Taking the flow parameter $\lambda = 1.8\, \fm^{-1}$, this model,
without an explicit 3NF, reproduces quite 
well not only the $\Hh$ binding energy but also the cross
section for $n$-$\Hh$ scattering in the energy regime 
with pronounced four-nucleon resonances
\cite{deltuva:08b}. Thus, this particular SRG potential
 yields a better description of the $3n$ + proton system  
and may be expected to  provide more solid conclusions
about the $3n$ resonances as compared to  other force
models. 

 $3n$ transition matrix elements calculated
with the above interactions
show no indications of resonances. A common procedure 
is to vary the strength of the potential to generate
an artificial resonance (or even a bound state)
and to follow its evolution towards physical strength
\cite{PhysRevC.66.054001,lazauskas:3n,PhysRevLett.118.232501,PhysRevLett.119.032501}. However, this way one may create also bound dineutron
states, thereby introducing additional thresholds
in the $3n$ system that complicate the analysis. In fact,
a bound  ${}^1S_0$ dineutron appears already for the enhancement
factor below 1.1 \cite{lazauskas:3n}.  For this reason and since the
${}^1S_0$ features are known quite well, the original potential
is kept in the ${}^1S_0$ partial wave while enhancing the potential
strength in all higher waves by the same factor $f$, i.e.,
\begin{equation} \label{eq:vf}
\langle p' L'sj| v |p Lsj \rangle
 = \langle p' L'sj| v_{nn} |p Lsj \rangle 
( \delta_{L'0}\delta_{L0} + f\bar{\delta}_{L'0} \bar{\delta}_{L0})
\end{equation}
where $v_{nn}$ is the physical two-neutron potential.
Bound dineutron appears in ${}^3P_2$ - ${}^3F_2$ 
(${}^3P_0$) partial wave at $f=7.24$ ($f=7.97$) when
using the SRG model but at $f=4.0$ ($f=5.95$) when
using the Reid93 model, the latter values being fully consistent with 
 Ref.~\cite{lazauskas:3n}. Although critical enhancement
factors depend quite strongly on the potential model,
the quantitative behavior of dineutron resonances, i.e., their
trajectories in the complex energy plane, are quite similar:
reducing $f$ they acquire large width and
for $f=1$ move deeply into the third quadrant becoming
 physically unobservable \cite{lazauskas:3n}.
Such a behavior of dineutron resonances is fully consistent
with Ref.~\cite{lazauskas:3n} and is therefore not shown here.

\begin{figure}[!]
\begin{center}
\includegraphics[scale=0.66]{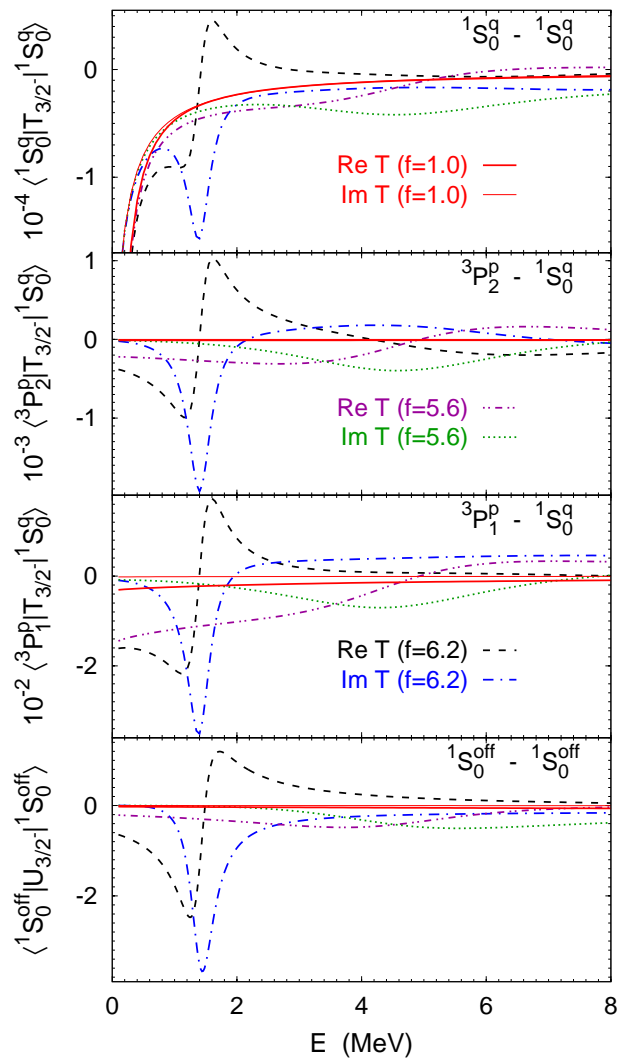}
\end{center}
\caption{\label{fig:treim} (Color online)
Energy dependence of real and imaginary parts
of  selected $J^\Pi = \frac32^-$
three-neutron transition matrix elements
calculated using SRG potential with higher wave
 enhancement factors $f=1.0$, 5.6 and 6.2.
For the off-shell state 
$\epsilon_{\rm off} = 9$ MeV was chosen.
Matrix elements are given in arbitrary units but preserving
the relative scale. }
\end{figure}

I start a detailed $3n$ system study with the  $J^\Pi = \frac32^-$
state that was predicted in Ref.~ \cite{PhysRevLett.118.232501} to
exhibit a resonance and  in Ref.~\cite{lazauskas:3n} 
 to be the most favorable for the existence of bound trineutron
when enhancing the $nn$ interaction in the single ${}^3P_2$ - ${}^3F_2$ 
 partial wave. In present work, unless explicitly stated otherwise, 
the $nn$ interaction is enhanced in all partial waves with $L \ge 1$
as given by Eq.~(\ref{eq:vf}).
The SRG model produces a bound trineutron when
$f=6.42$. For lower $f$ values a resonant behavior of $T$ and $U$ operators
 can be seen as demonstrated in Fig.~\ref{fig:treim}
for several choices of initial and final channels of all three types
described in previous section.
If the two-neutron  ${}^1S_0$ state with vanishing relative momentum
$p=0$  is interpreted as an unbound dineutron, 
the matrix element $\langle {}^1S_0^q | T_{J^\Pi} | {}^1S_0^q \rangle $
can be interpreted as the amplitude for ``elastic'' 
neutron-dineutron scattering.
When the $3n$ system energy approaches zero, this amplitude
diverges as the driving term in Eq.~(\ref{eq:AGSt}) does
due to kinematic reasons discussed in the previous section.  This is not
a resonance and it is not seen in other  matrix elements
of Fig.~\ref{fig:treim}
where  $\langle {}^3P_j^p | T_{J^\Pi} | {}^1S_0^q \rangle $
can be interpreted as the neutron-dineutron ``breakup'' amplitude
in collinear kinematics \cite{gloeckle:96a,deltuva:03a}.
The off-shell matrix element
$\langle {}^1S_0^{\rm off} | U_{J^\Pi} | {}^1S_0^{\rm off} \rangle $
 shown in last panel of Fig.~\ref{fig:treim} 
has no direct physics interpretation. The most important message
is that, if a given Hamiltonian supports a resonance,
   all matrix elements, despite their differences by several
orders of magnitude or the repulsive character of the
final-state interaction as in the ${}^3P_1$ wave, as functions of energy
 exhibit resonant behavior corresponding to the same 
(within numerical accuracy) values $E_r -i\Gamma/2$.
This is most evident for $f=6.2$ where the corresponding resonance 
at $(1.41-0.22i)$ MeV is most pronounced. 
Decreasing $f$ the pole moves to higher energy and away from the real axis; 
as a consequence, at $f=5.6$ with  $E_r -i\Gamma/2 = (4.36 - 2.24i)$ MeV
the resonant behavior is far less pronounced.
For the physical interaction strength $f=1$ also shown in Fig.~\ref{fig:treim} 
by solid curves
the resonant behavior can not be seen, and the magnitude
 of matrix elements is significantly smaller.

\begin{figure}[!]
\begin{center}
\includegraphics[scale=0.64]{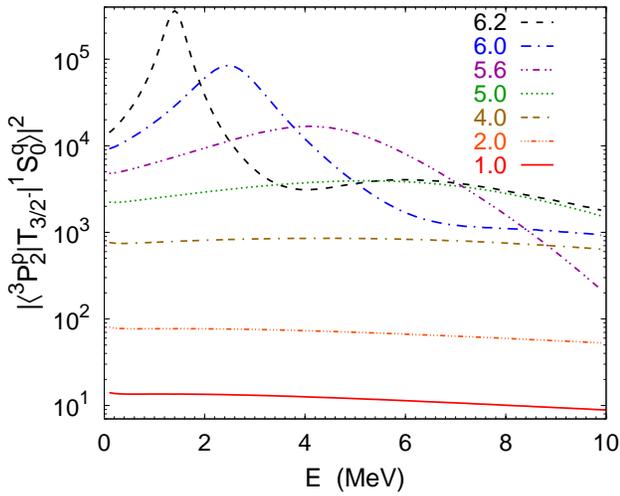}
\end{center}
\caption{\label{fig:t2-3p2} (Color online)
Energy dependence of transition strengths 
$|\langle {}^3P_2^p | T_{3/2^-} | {}^1S_0^q \rangle|^2 $ 
obtained using the SRG potential with higher-wave
 enhancement factors $f=6.2$, 6.0, 5.6, 5.0, 4.0, 2.0, and 1.0.
Transition strengths are given in arbitrary units but preserving
the relative scale. }
\end{figure}

Transition strengths (probabilities) are proportional to squares of 
amplitudes; an example is shown in Fig.~\ref{fig:t2-3p2} 
for ``breakup'' transitions
for the potential enhancement factor $f$ ranging
from 6.2 to 1.
As expected, with decreasing $f$ resonant peaks move to higher
energy and become wider, disappearing around $f=4.0$, i.e., well above
the physical interaction strength $f=1$.
This fact strongly suggests the absence of physically observable
$3n$ resonance in the  $J^\Pi = \frac32^-$ state,
confirming the conclusions of Refs.~\cite{PhysRevC.66.054001,lazauskas:3n}.

\begin{figure}[!]
\begin{center}
\includegraphics[scale=0.64]{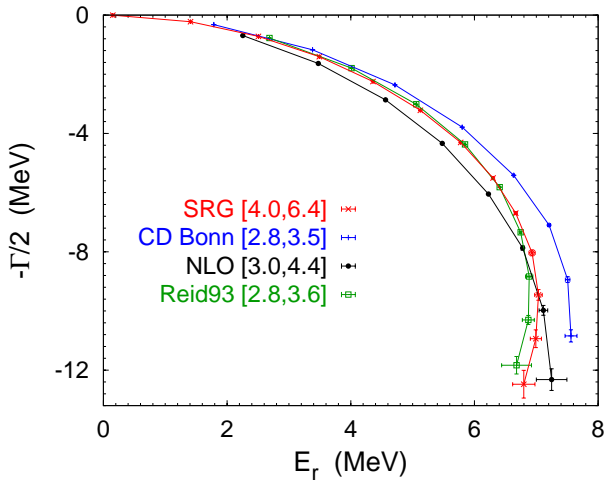}
\end{center}
\caption{\label{fig:res32-} (Color online)
Three-neutron  $J^\Pi =\frac32^-$ resonance trajectories for SRG, CD Bonn, NLO,
and Reid93 potentials obtained varying the higher-wave
(only ${}^3P_2$ - ${}^3F_2$ for Reid93)
enhancement factor $f$ in the given interval with
the step of 0.1 (CD Bonn and Reid93) or 0.2 (SRG and NLO).
Lines are for guiding the eye only.}
\end{figure}

Further support for the above conclusion comes from the
Fig.~\ref{fig:res32-} where the extracted $J^\Pi =\frac32^-$ 
transition operator pole trajectories  in the complex energy plane
are shown not only for SRG but also for CD Bonn, NLO, and Reid93 potentials.
In the latter case, for the comparison with Ref.~\cite{lazauskas:3n},
the potential was enhanced in the ${}^3P_2$ - ${}^3F_2$ wave only.
The pole trajectory for Reid93 in Fig.~\ref{fig:res32-}
is in good agreement with the results of Ref.~\cite{lazauskas:3n}.
For soft potentials like  NLO and, especially, SRG, 
the evolution with $f$ is slower than for CD Bonn and Reid93.
As a consequence, NLO and and SRG need larger 
$f$ values to exhibit a $3n$ resonance. Apart from that
the trajectories are qualitatively similar for all potential models:
decreasing the enhancement factor $f$ the pole moves to higher energy
and away from the real axis until the turning point
$E_r - i\Gamma/2 \approx (7- 10i)$ MeV where  $E_r$ starts to decrease
while $\Gamma$ is rapidly increasing. Still, at this point
$f$ is around 3 or 4, indicating that the model system is far
from the physical one with $f=1$.
Decreasing $f$ below 3 or 4, depending on the potential, 
 the pole moves too far from the real axis 
to be seen as a scattering resonance and its position therefore 
cannot be reliably
extracted from $3n$ transition operators calculated on the real axis.
This fact is reflected in increased theoretical error bars,
estimated from calculations with different initial and final states
and with different number of terms (typically, $n \le 2$ to 4) 
in Eq.~(\ref{eq:Upole}). 

\begin{figure}[!]
\begin{center}
\includegraphics[scale=0.69]{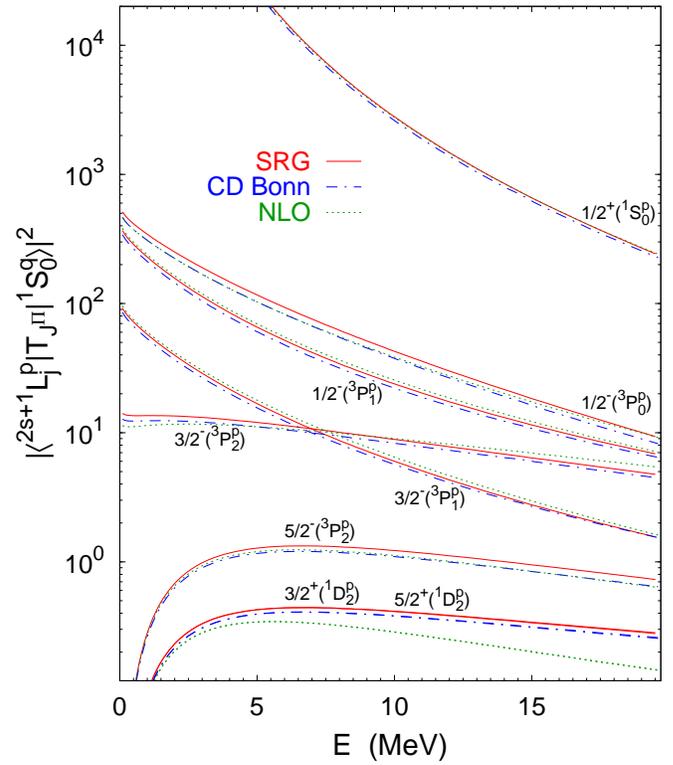}
\end{center}
\caption{\label{fig:t2j} (Color online)
Energy dependence of transition strengths 
$|\langle {}^{2s+1}L_j^p | T_{J^\Pi} | {}^1S_0^q \rangle|^2 $ 
for $J \le \frac52$ states
obtained using  physical SRG, CD Bonn, and NLO potentials.
Transition strengths are given in arbitrary units but preserving
the relative scale. 
$\frac32^+$ and $\frac52^+$ results are indistinguishable in the plot.
}
\end{figure}

\begin{figure}[!]
\begin{center}
\includegraphics[scale=0.64]{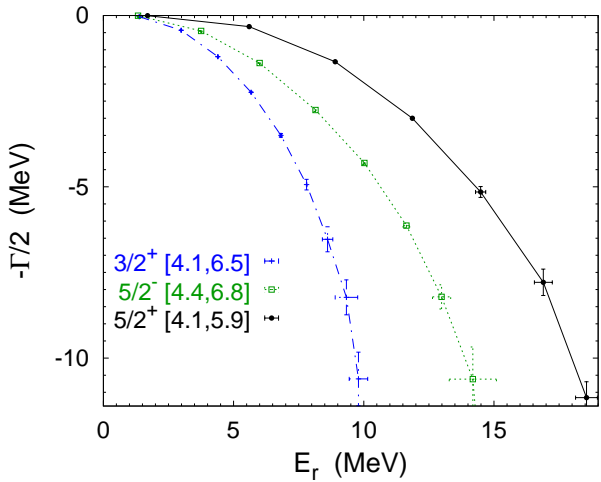}
\end{center}
\caption{\label{fig:resj} (Color online)
Three-neutron  $J^\Pi =\frac32^+$, $\frac52^-$, and  $\frac52^+$
 resonance trajectories for the SRG potential obtained varying 
the higher-wave enhancement factor $f$ in the given interval with
the step of 0.3. Lines are for guiding the eye only. }
\end{figure}

For the physical interaction strength $f=1$ no resonant behavior
is seen also in other $J^\Pi$ states. This is illustrated
in Fig.~\ref{fig:t2j} taking as example ``breakup'' transition 
strengths 
$|\langle {}^{2s+1}L_j^p | T_{J^\Pi} | {}^1S_0^q \rangle|^2 $
for all $J \le \frac52$. The results obtained with
SRG, CD Bonn, and NLO potentials show some model dependence but 
all are consistent with the absence of an observable $3n$ 
resonance. The strongest transition strength seen in the
 $J^\Pi =\frac12^+$ state is mostly due to the
 final-state $nn$ $t$-matrix that acts in the ${}^1S_0$ wave 
as compared to weaker $P$ or $D$ waves for other $J^\Pi$ states.
For vanishing energy this amplitude diverges 
as the corresponding  driving term in Eq.~(\ref{eq:AGSt}) does,
but it is not really resonant. In fact, in the considered
$nn$ interaction enhancement scheme the
$J^\Pi =\frac12^{\pm}$ states
are even less favorable for trineutron resonances
than $J^\Pi =\frac32^{\pm}$ and  $J^\Pi =\frac52^{\pm}$ states:
At $f=7.24$, with the bound ${}^3P_2$ - ${}^3F_2$ 
dineutron appearing in the SRG model, $J^\Pi =\frac12^{\pm}$
trineutrons are still not bound. In contrast, trineutrons
in $J^\Pi =\frac32^{+}$, $\frac52^{+}$, and $\frac52^{-}$
states become bound at $f=6.71$, 6.02, and 6.94, respectively.
Their pole trajectories when reducing the enhancement
factor $f$ are shown in Fig.~\ref{fig:resj}. Qualitatively,
the behavior is similar to  the $J^\Pi =\frac32^{-}$
case of Fig.~\ref{fig:res32-}, but the real part $E_r$ 
reaches higher values, especially for $J^\Pi =\frac52^{+}$.
Again, decreasing $f$ below 4, 
 the pole moves far away from the real axis and
does not manifest itself as a visible  scattering resonance.

\begin{figure}[!]
\begin{center}
\includegraphics[scale=0.64]{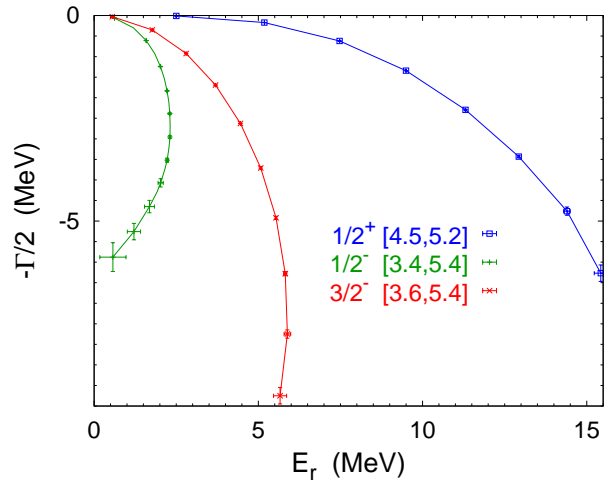}
\end{center}
\caption{\label{fig:resj1} (Color online)
Three-neutron  $J^\Pi =\frac12^+$, $\frac12^-$, and  $\frac32^-$
 resonance trajectories obtained  with the 
attractive ${}^3P_1$ potential as described in the text.
Results are based on the SRG model while
the enhancement factor $f$ is varied in the given interval with
the step of 0.1 (0.2) for positive (negative) parity states. }
\end{figure}

Of course, resonance trajectories depend on the interaction enhancement scheme
but the physical point $f=1$ is the same. In order to avoid the
presence of bound dineutron,
a different $nn$ interaction enhancement scheme is used for the
study of $J^\Pi =\frac12^{\pm}$ resonance trajectories: In the 
originally repulsive ${}^3P_1$  partial wave the factor $f$ in Eq.~(\ref{eq:vf})
is replaced by $(2-f)$ such that  $f=1$ as before corresponds to the physical
strength but the ${}^3P_1$ potential becomes attractive with increasing $f$.
In other waves the potential (\ref{eq:vf}) is used.
Using the SRG model in this scheme ${}^3P_1$ dineutron becomes bound at $f=6.47$ while
$J^\Pi =\frac12^{+}$, $\frac12^{-}$, and $\frac32^{-}$ trineutrons become bound at
$f=5.28$, 5.48,  and 5.44, respectively. Their resonance trajectories
are shown in Fig.~\ref{fig:resj1}. Trajectories for $J^\Pi =\frac12^{+}$
and  $\frac32^{-}$ are similar to those in 
Figs.~\ref{fig:res32-} and \ref{fig:resj} while
the $\frac12^{-}$ trajectory stays much closer to the imaginary axis, i.e.,
the turning point is around $E_r - i\Gamma/2 \approx (2.3- 2.7i)$ MeV. 
Thus, the $\frac12^{-}$ resonance most evidently exhibits the 
trend to move to the $E_r < 0$ region for $f=1$,
becoming physically unobservable.

An alternative approach to generate an artificial $3n$ bound state or resonance 
is by adding an attractive 3NF; only the total
isospin $\frac32$ component is acting in the $3n$ system.
However, the contribution of a realistic 3NF,
being of short range, is  suppressed by the Pauli repulsion 
\cite{PhysRevC.93.044004,lazauskas:3n}, and unphysically 
strong 3NF is needed to achieve a visible effect. 
Consistently with this observation, Ref.~\cite{PhysRevLett.118.232501}
found the effect of a realistic
 $\chi$EFT 3NF in few-neutron systems to be very small.
This suggests that the absence of an explicit 3NF does not affect
conclusions on the absence of the resonant behavior.

\section{Conclusions \label{sec:sum}}

The three-neutron system was studied using exact Faddeev-type equations
for transition operators that were solved numerically in the
momentum-space framework. Various on-shell and off-shell
matrix elements were calculated  searching for their poles
in the complex energy plain leading to resonant behavior.
An important advantage of the present transition operator approach 
as compared to previous bound-state-type studies
is its ability to estimate not only the 
resonance position but also its effect on scattering amplitudes
that include both resonant (if present) and nonresonant
(also called background) contributions and their interference
in collision observables. Since  $3n$ elastic scattering 
experiments are so far technically impossible,  this work restricted
itself to few selected transition strengths related to
the $3n$ collision process; this was  sufficient to draw
conclusions on $3n$ resonances.

All tested physical $nn$ force models,
including the SRG potential successfully reproducing the
resonant $n$-$\Hh$ cross section, 
 were found to exclude the possibility of an observable
$3n$ resonance. To generate artificial $3n$ resonances 
(or even bound states) the 
$nn$ interaction was enhanced in higher $nn$ partial waves while
keeping the original physical strength in the  ${}^1S_0$ partial wave.
For appropriate (state $J^\Pi$ and  potential-dependent) 
enhancement factor values the
resonant behavior was observed in all studied 
 on-shell and off-shell matrix elements of $3n$ transition operators.
However, the resonant behavior disappears with the
enhancement factor $f$ still having a value around 3 or 4, i.e.,
for systems that cannot be considered as realistic
$3n$ systems. In these situations the resonance pole
is typically more than 10 MeV away from the real axis while
in transition amplitudes and strengths the background contribution 
dominates over the resonant one such that no resonant
behavior can be seen. For this reason the transition operator
pole trajectory in the
complex energy plane could not be reliably followed towards the
physical limit $f=1$, but it can be expected to be even further away 
from the real axis thereby excluding the physical observability
of the $3n$ resonance. This conclusion is fully consistent
with Refs.~\cite{PhysRevC.93.044004,PhysRevC.66.054001,lazauskas:3n} but
contradicts Ref.~\cite{PhysRevLett.118.232501}. The latter 
work, however, employed some questionable procedures such
as the trineutron bound-state calculation  above the
dineutron threshold or the extrapolation of energy into
a different sheet of the complex energy plane.

Despite the absence of resonant states, three-neutron transition
amplitudes depend on the available energy and final state kinematics;
thus, the existence of 
some peak structures in the transition strengths can not be excluded.
In fact, the $nn$ final-state interaction kinematics with vanishing
 two-neutron relative energy should correspond to a rather sharp 
peak due to the ${}^1S_0$ virtual state as in the
neutron-deuteron breakup \cite{gloeckle:96a,deltuva:03a}.

Given the controversy in the literature regarding the
four-neutron resonance 
\cite{PhysRevC.93.044004,PhysRevLett.117.182502,PhysRevLett.118.232501,PhysRevLett.119.032501},
the extension of the present transition operator study to the
$4n$ system is of high importance and interest. The work into this direction
is underway.

\vspace{1mm}

The author acknowledges discussions with R. Lazauskas,
the support  by the Alexander von Humboldt Foundation
under grant no. LTU-1185721-HFST-E, and
the hospitality of the Ruhr-Universit\"at Bochum
where a part of this work was performed.


\end{document}